\documentclass[conference,letterpaper]{IEEEtran}
\IEEEoverridecommandlockouts

\usepackage[letterpaper, left=0.78in, right=0.78in, top=1.02in, bottom=0.78in]{geometry}

\usepackage{cite}
\usepackage{amsmath,amssymb,amsfonts}
\usepackage[ruled,vlined,linesnumbered]{algorithm2e}
\usepackage{graphicx}
\usepackage{booktabs}
\usepackage{tikz}
\usepackage{xurl}

\begin{document}

\bstctlcite{IEEEexample:BSTcontrol}

\title{Adapting Dijkstra for Buffers and Unlimited Transfers \textsuperscript{*}\thanks{*Authors are listed in alphabetical order.}}

\author{
\IEEEauthorblockN{Denys Katkalo\IEEEauthorrefmark{1},
Andrii Rohovyi\IEEEauthorrefmark{2},
Toby Walsh\IEEEauthorrefmark{2}}
\IEEEauthorblockA{\IEEEauthorrefmark{1}Igor Sikorsky Kyiv Polytechnic Institute, Kyiv, Ukraine\\katkalo.denys@lll.kpi.ua}
\IEEEauthorblockA{\IEEEauthorrefmark{2}University of New South Wales, Sydney, Australia\\\{a.rohovyi, t.walsh\}@unsw.edu.au}
}

\maketitle

\begin{abstract}
In recent years, RAPTOR based algorithms have been considered the state-of-the-art for path-finding with unlimited transfers without preprocessing. However, this status largely stems from the evolution of routing research, where Dijkstra-based solutions were superseded by timetable-based algorithms without a systematic comparison. In this work, we revisit classical Dijkstra-based approaches for public transit routing with unlimited transfers and demonstrate that Time-Dependent Dijkstra (TD-Dijkstra) outperforms MR. However, efficient TD-Dijkstra implementations rely on filtering dominated connections during preprocessing, which assumes passengers can always switch to a faster connection. We show that this filtering is unsound when stops have buffer times, as it cannot distinguish between seated passengers who may continue without waiting and transferring passengers who must respect the buffer. To address this limitation, we introduce Transfer Aware Dijkstra (TAD), a modification that scans entire trip sequences rather than individual edges, correctly handling buffer times while maintaining performance advantages over MR. Our experiments on the London and Switzerland networks show that we can achieve more than a twofold speedup over MR while producing optimal results on both networks, with and without buffer times.
\end{abstract}


\section{Introduction}

In this work, we revisit the classical Dijkstra algorithm and observe that it outperforms standard unlimited transfer algorithms such as MR (introduced as MR-inf in~\cite{delling2013computing} and renamed in~\cite{baum2019ultra}). We also adapt it to correctly handle buffer times, making it suitable for realistic public transit routing. We call this adaptation \emph{Transfer Aware Dijkstra (TAD)}. The key insight is that standard Time-Dependent Dijkstra relies on dominated connection filtering, which fails when buffer times are present because it cannot distinguish between seated passengers (who may continue without waiting) and transferring passengers (who must respect the buffer). TAD addresses this by scanning entire trip sequences rather than individual edges, correctly modeling the asymmetry between seated and transferring passengers while maintaining the performance advantages of Dijkstra-based approaches.

The development of public transit routing algorithms has progressed \cite{car_vs_transport} from classical Time-Dependent (TD) Dijkstra ~\cite{dijkstra1959note, multi_dijkstra} to timetable-based techniques such as RAPTOR~\cite{raptor}, CSA~\cite{csa}, Trip-based routing~\cite{trip_based_routing, trip_based_routing_condense}, and Transfer Patterns~\cite{transfer_patterns}. RAPTOR marked a major shift, replacing TD-Dijkstra in many practical applications including Bing Maps, OpenTripPlanner, R5, Navitia.io, and Solari. During this evolution, Dijkstra-based techniques received less attention and were not thoroughly evaluated in modern multimodal scenarios.

For the unlimited transfer problem, where passengers may walk, cycle, or use other non-scheduled modes, RAPTOR-based algorithms such as MR and MCR~\cite{baum2019ultra, delling2013computing} have been treated as state-of-the-art. However, this assumption is based on the historical evolution of routing research rather than a systematic comparison. A previous study~\cite{rohovyi2025multimodal}\nocite{rohovyyi2024ttn} demonstrated that TD-Dijkstra can outperform CSA and RAPTOR on unlimited transfer problems, but did not compare against MR. We extend that analysis and demonstrate that TD-Dijkstra variants outperform MR in this setting.

While TD-Dijkstra shows strong performance, efficient implementations rely on filtering dominated connections during preprocessing. A connection $(d_1, a_1)$ dominates $(d_2, a_2)$ if $d_1 \geq d_2$ and $a_1 \leq a_2$, meaning it departs no earlier and arrives no later. This filtering assumes passengers can always switch to a dominating connection without penalty.

However, this assumption fails when stops have buffer times. The GTFS specification allows each stop to define a buffer time that passengers must wait before boarding a connecting service. Crucially, this buffer applies only to transferring passengers; a seated passenger continuing on the same trip incurs no delay. Dominated connection filtering cannot capture this distinction because it analyzes edges in isolation, without knowledge of whether the passenger will remain seated or transfer. TAD overcomes this limitation while preserving the performance advantages of Dijkstra-based routing.

\section{Preliminaries}
This section defines key terminology and presents the foundational algorithms.

\subsection{Terminology}

\subsubsection*{Network}
We define a public transit network as a 4-tuple $(\mathcal{S}, \mathcal{T}, \mathcal{R}, G)$, where $\mathcal{S}$ denotes the set of stops, $\mathcal{T}$ the set of trips, $\mathcal{R}$ the set of routes, and $G = (V, E)$ a directed, weighted transfer graph. A stop represents any location within the network where passengers may enter or exit a vehicle, encompassing various transit modes including buses, trains, and ferries. A route $R \in \mathcal{R}$ groups trips that share the same ordered stop pattern and can therefore be scanned in a common stop order. Each trip $T \in \mathcal{T}$ consists of an ordered sequence of stop events $\langle \varepsilon_0, \ldots, \varepsilon_k \rangle$ executed by a single vehicle. A stop event $\varepsilon$ is characterized by a triple $(\tau_{\text{arr}}(\varepsilon), \tau_{\text{dep}}(\varepsilon), v(\varepsilon))$, where $v(\varepsilon)$ identifies the stop, $\tau_{\text{arr}}(\varepsilon)$ specifies the vehicle's arrival time, and $\tau_{\text{dep}}(\varepsilon)$ indicates its subsequent departure time. We use $T[i]$ to reference the $i$-th stop event within trip $T$, and define the trip length as $|T| := k$, representing the total number of stop events.

The transfer graph $G = (V, E)$ comprises a vertex set $V$ that contains all stops ($\mathcal{S} \subseteq V$) and an edge set $E \subseteq V \times V$. Each edge $e = (v, w) \in E$ is associated with a transfer time $\tau_{\text{tra}}(e)$. For any path $P = \langle v_1, \ldots, v_k \rangle$ in $G$, we extend this notion by defining $\tau_{\text{tra}}(P) := \sum_{i=1}^{k-1} \tau_{\text{tra}}((v_i, v_{i+1}))$. Our algorithm places no constraints on the structure of $G$ and does not require precomputation: transitive closure is not required, strong connectivity is permitted, and transfer times may correspond to walking, cycling, or any other mode of travel that operates independently of fixed schedules.

Timetable-based algorithms such as CSA and RAPTOR do not process transfer connections incrementally during their main loop. Instead, they require the transfer graph to be \emph{transitively closed}. A graph $G = (V, E)$ is transitively closed if, whenever there exists a path from vertex $a$ to vertex $c$, there is also a direct edge $(a, c) \in E$. Formally, if $(a, b) \in E$ and $(b, c) \in E$, then $(a, c) \in E$ must hold. In practice, transitively closed transfer graphs are constructed by inserting direct edges between all stop pairs whose shortest path distance falls below a specified threshold (e.g., 5--10 minutes of walking).

However, for the unlimited transfer problem, where the transfer graph may span an entire road network with millions of vertices, computing the transitive closure becomes computationally infeasible. Limiting the maximal transfer duration to 20 minutes before computing the transitive closure already leads to a graph that is too large for practical applications~\cite{unrestricted_walking}. This limitation motivates preprocessing techniques such as ULTRA, which avoid the need for transitive closure by precomputing only the transfer shortcuts that are actually required for optimal journeys.

\subsubsection*{Buffer Times}

The GTFS specification defines minimum transfer times between pairs of stops in the optional \texttt{transfers.txt} file. When the source and destination stops are identical, this represents the interchange time within a station: the minimum time a passenger must wait after alighting before boarding a connecting service at the same stop. We denote this buffer time as $\beta(v)$ for stop $v \in \mathcal{S}$. It models real-world constraints such as platform changes, walking between arrival and departure points within a station, or safety margins for schedule variance.

Critically, buffer times apply only to transferring passengers. A passenger who remains seated on the same trip does not incur the buffer time, even if the vehicle stops at that location. This asymmetry between seated and transferring passengers has important implications for routing algorithms.

\subsubsection*{Journeys}
A journey characterizes a passenger's movement through the network from a source vertex $s \in V$ to a target vertex $t \in V$. Each segment where the passenger rides a public transit vehicle is represented by a trip segment, while transitions between rides are modeled as paths in the transfer graph.

A journey $J = \langle P_0, T_0^{ij}, \ldots, T_{k-1}^{mn}, P_k \rangle$ is an alternating sequence of transfers and trip segments, where transfers may be empty (consisting of a single stop). For source and target vertices $s, t \in V$, we call $J$ an \emph{$s$-$t$-journey} if $P_0$ begins at $s$ and $P_k$ concludes at $t$. The departure time is $\tau_{\text{dep}}(J) := \tau_{\text{dep}}(T_0[i]) - \tau_{\text{tra}}(P_0)$, and the arrival time is $\tau_{\text{arr}}(J) := \tau_{\text{arr}}(T_{k-1}[n]) + \tau_{\text{tra}}(P_k)$. The number of trips is denoted $|J| := k$. A special case arises when $J = \langle P_0 \rangle$ consists solely of a transfer graph path; since no trips are involved, $\tau_{\text{dep}}(J)$ must be specified separately, with $\tau_{\text{arr}}(J) := \tau_{\text{dep}}(J) + \tau_{\text{tra}}(P_0)$.

When computing boarding times for transfers, the buffer time must be respected. If a passenger arrives at stop $v$ at time $\tau$ via a transfer (not by remaining seated), the earliest time they can board a departing trip is $\tau + \beta(v)$.

\subsubsection*{Non-FIFO Schedules}

A set of trips on the same edge may violate the FIFO property: a later-departing trip arrives earlier than an earlier-departing trip. Consider two trips on edge $A \to B$:
\begin{align*}
T_1&: A \xrightarrow{\text{dep}=\text{08:00}} B\ (\text{arr}=\text{09:30}) \\
T_2&: A \xrightarrow{\text{dep}=\text{08:30}} B\ (\text{arr}=\text{09:00})
\end{align*}
Here $T_2$ departs 30 minutes later but arrives 30 minutes earlier than $T_1$. This commonly occurs when express and local services share stops, or when different transportation modes (e.g., bus vs.\ train) serve the same origin-destination pair. A naive algorithm that only considers the first available trip would miss the faster option.

\subsubsection*{Filtering Dominated Connections}

Non-FIFO schedules can be resolved by filtering dominated connections during preprocessing. A connection $(d_1, a_1)$ \emph{dominates} another connection $(d_2, a_2)$ on the same edge if $d_1 \geq d_2$ and $a_1 \leq a_2$, with at least one inequality strict. In the example above, $T_2$ dominates $T_1$ since $T_2$ departs later but arrives earlier. After filtering, only non-dominated connections remain, restoring the FIFO property and enabling standard TD-Dijkstra.

This filtering approach is sound when no buffer times exist: a passenger who boards a dominated connection could always switch to the dominating connection and arrive no later. However, as we show next, this reasoning fails when buffer times are present.

\subsubsection*{Why Filtering Fails with Buffer Times}

When stops have buffer times, filtering dominated connections may exclude optimal journeys. The fundamental issue is that dominated connection filtering analyzes edges in isolation, without distinguishing between seated passengers (who incur no buffer) and transferring passengers (who must wait).

\paragraph{Motivating Example.}

Consider stops $A$, $B$, $C$ with buffer $\beta(B) = 20$ min. Trip $T_1: A \xrightarrow{8:00} B \xrightarrow{9:40} C$ (arrives 10:30). Trip $T_2: A \xrightarrow{8:30} B$ (arrives 9:30, terminates).

On edge $A \to B$, we have two connections:
\begin{itemize}
    \item Connection from $T_1$: departs 08:00, arrives 09:40
    \item Connection from $T_2$: departs 08:30, arrives 09:30
\end{itemize}

The connection from $T_2$ dominates the connection from $T_1$: it departs later (08:30 $>$ 08:00) and arrives earlier (09:30 $<$ 09:40). Standard preprocessing would filter out $T_1$'s connection on edge $A \to B$.

Now consider a query from $A$ to $C$ departing at 07:50:

\textbf{Optimal journey (without filtering):} Board $T_1$ at $A$ (depart 08:00), remain seated through $B$, arrive at $C$ at 10:30. The passenger stays on $T_1$ continuously and does not need to wait for the buffer time at $B$.

\textbf{Journey found by TD-Dijkstra (with filtering):} The dominated connection from $T_1$ has been removed. Board $T_2$ at $A$ (depart 08:30), arrive at $B$ at 09:30. To board another trip, the passenger must wait for the buffer time: earliest boarding time is $09:30 + 20 = 09:50$. But $T_1$'s connection from $B$ to $C$ departs at 09:40, so the passenger misses this connection and must wait for a later service.

The filtered algorithm produces a suboptimal result because it cannot represent the fact that a seated passenger on $T_1$ may continue to $C$ without waiting, while a transferring passenger arriving on $T_2$ must respect the 20-minute buffer.

\paragraph{General Principle.}
Dominated connection filtering fails with buffer times because:
\begin{enumerate}
    \item It analyzes each edge independently, without knowledge of whether the passenger will continue on the same trip or transfer.
    \item A ``dominated'' connection may be part of a longer trip that allows the passenger to bypass buffer times by remaining seated.
    \item The dominating connection, while faster to a single stop, may force a transfer that incurs buffer time, ultimately resulting in a later arrival.
\end{enumerate}

This demonstrates that dominated connection filtering is unsound when buffer times are present in the timetable. Consequently, TD-Dijkstra with dominated connection filtering cannot be applied to networks with buffer times.

\subsection{Algorithms}

\subsubsection*{Dijkstra's Algorithm}

Given a graph $G = (V, E)$ with non-negative edge lengths $\ell : E \to \mathbb{R}^+_0$ and a source vertex $s \in V$, Dijkstra's algorithm computes the shortest path length from $s$ to every vertex $v \in V$. The algorithm maintains a tentative distance $\text{dist}[v]$ for each vertex, initialized to $\infty$, along with a priority queue $Q$ that orders vertices by their tentative distances. The source $s$ is inserted into $Q$ with $\text{dist}[s] = 0$. Vertices are then extracted from $Q$ in order of increasing distance. When a vertex $v$ is extracted, it is settled by relaxing all outgoing edges. Relaxing an edge $e = (v, w) \in E$ compares $\text{dist}[w]$ against $\text{dist}[v] + \ell(e)$; if the latter is smaller, $\text{dist}[w]$ is updated and $w$ is inserted into $Q$ with the new distance as its key.

\subsubsection*{Time-Dependent Dijkstra}

Time-Dependent Dijkstra (TD-Dijkstra)~\cite{Pyrga2008Efficient} extends classical Dijkstra's algorithm to handle graphs where edge weights vary with time. In public transit networks, travel times depend on departure time due to scheduled vehicle departures. Rather than fixed edge lengths, TD-Dijkstra operates on a time-dependent graph where each edge $e = (v, w)$ has an associated arrival time function $\ell_e(\tau) : \mathbb{R}^+_0 \to \mathbb{R}^+_0$ that maps departure time $\tau$ to the arrival time when traversing $e$.

The algorithm maintains tentative arrival times $\tau_{\text{arr}}[v]$ instead of distances. When relaxing an edge $e = (v, w)$, the arrival time at $w$ is computed as $\ell_e(\tau_{\text{arr}}[v])$, reflecting that the travel time depends on when the traveler departs from $v$. Efficient implementations filter dominated connections during preprocessing to enable binary search, but as shown above, this approach is invalid when buffer times are present.

Prior work~\cite{rohovyi2025multimodal} has shown that TD-Dijkstra outperforms CSA and RAPTOR on the unlimited transfer problem because it does not require computing the transitive closure of the transfer graph. That work also introduced Timetable Nodes (TTN)~\cite{rohovyi2025multimodal, rohovyyi2024ttn}, a technique that aggregates departure times at the node level and uses data structures such as Fractional Cascading to reduce the number of binary searches from $k$ (one per outgoing edge) to one. While TTN provides significant speedups in the original Python implementation, we found that all TTN variants are substantially slower than the classic per-edge binary search approach in our C++ implementation (28--142\% slower depending on variant and CH type; see Table~\ref{tab:ttn_comparison}). We attribute this to two factors: first, timetable data stored in contiguous \texttt{std::vector} structures already exhibits excellent cache locality, and modern CPU prefetching effectively hides the latency of sequential binary searches; second, the auxiliary data structures required by TTN variants (fractional cascading pointers, combined search trees, balanced trees) introduce additional memory indirection and cache pressure that outweigh the reduction in binary searches.

In our experiments, we run TD-Dijkstra on the core graph and test Bucket-CH acceleration for the first and last transfers.

\subsubsection*{Contraction Hierarchies}

Contraction Hierarchies (CH)~\cite{geisberger2012ch} is a preprocessing technique for accelerating shortest path queries in graphs. The fundamental operation is vertex contraction: removing a vertex from the graph while inserting shortcut edges between its neighbors to preserve shortest path distances. During preprocessing, vertices of a graph $G = (V, E)$ are contracted in a heuristically determined order, with each vertex's position in this order defining its rank. The result is an augmented graph $G^+ = (V, E^+)$ containing both original and shortcut edges. This augmented graph decomposes into an upward graph $G^{\uparrow} = (V, E^{\uparrow})$ with edges directed from lower-ranked to higher-ranked vertices, and a downward graph $G^{\downarrow} = (V, E^{\downarrow})$ with the reverse orientation. Queries are answered using bidirectional Dijkstra, with the forward search exploring $G^{\uparrow}$ and the backward search exploring $G^{\downarrow}$.

\subsubsection*{Core-CH}

Core-CH~\cite{bauer2010combining, baum2019ultra, delling2013computing} is a variant of Contraction Hierarchies designed for multimodal route planning. Unlike standard CH, Core-CH prohibits the contraction of vertices that correspond to stops, leaving a set of core vertices $V_c$ with $\mathcal{S} \subseteq V_c \subseteq V$ uncontracted. In addition to the partially augmented graph, this produces a core graph $G_c = (V_c, E_c)$ consisting of the core vertices and all shortcuts inserted between them. If only stops were permitted as core vertices, the number of core edges would grow quadratically with the number of stops, rendering both preprocessing and queries impractical. To address this, the contraction process is terminated once the average vertex degree in the core graph exceeds a specified threshold.

\subsubsection*{Bucket-CH}

Bucket-CH~\cite{knopp2007manytoMany, geisberger2012ch} extends Contraction Hierarchies to efficiently handle one-to-many queries. The algorithm proceeds in three phases. First, standard CH preprocessing is performed on the graph $G = (V, E)$. Second, given a set of target vertices $V_t \subseteq V$, a bucket storing distances to targets is computed for each vertex. This is achieved by running a backward search on $G^{\downarrow}$ from every target $t \in V_t$; for each vertex $v$ settled at distance $\text{dist}(v, t)$, the entry $(t, \text{dist}(v, t))$ is added to the bucket of $v$. Third, given a source vertex $s$, a forward search is performed on $G^{\uparrow}$. For each settled vertex $v$ at distance $\text{dist}(s, v)$, the algorithm scans the bucket of $v$: for each entry $(t, \text{dist}(v, t))$, the current shortest distance to $t$ is compared against $\text{dist}(s, v) + \text{dist}(v, t)$ and updated if improved.

\subsubsection*{RAPTOR}

RAPTOR answers single-source earliest-arrival queries to either a fixed target or to all stops simultaneously. In contrast to TD-Dijkstra, it requires the stop-to-stop transfer graph of the public transit network to be transitively closed. The algorithm operates in rounds, where round $i$ discovers journeys with exactly $i$ trips by extending journeys from round $i-1$. For each stop $v \in \mathcal{S}$ and round $i$, the arrival time $\tau_{\text{arr}}(v, i)$ represents the earliest known arrival at $v$ using at most $i$ trips. A query from source $s$ with departure time $\tau_{\text{dep}}$ initializes $\tau_{\text{arr}}(s, 0) = \tau_{\text{dep}}$ and all other arrival times to $\infty$, then iterates until no improvements occur.

Each round consists of two phases. In the route scanning phase, the algorithm sequentially scans all routes passing through stops improved in the previous round, tracking the earliest boardable trip $T_{\min}$. At each stop $v$ along the route, it checks whether alighting from $T_{\min}$ improves $\tau_{\text{arr}}(v, i)$, and whether an earlier trip becomes boardable. In the transfer relaxation phase, the algorithm relaxes all outgoing edges from improved stops, updating $\tau_{\text{arr}}(w, i)$ whenever $\tau_{\text{arr}}(v, i) + \tau_{\text{tra}}(e) < \tau_{\text{arr}}(w, i)$ for edge $e = (v, w)$.

RAPTOR correctly handles buffer times by applying them only when boarding a new trip, not when continuing on the same trip. This is possible because RAPTOR explicitly tracks trip membership during route scanning.

\subsubsection*{MR}

MR is a bicriteria route planning algorithm that extends the RAPTOR framework to support multimodal scenarios with unlimited transfers. It leverages a core graph computed via Core-CH to accelerate transfer relaxation.

Unlike standard RAPTOR, MR maintains tentative arrival times $\tau_{\text{arr}}(v, i)$ for every core vertex $v \in V_c$, not just stops. The transfer relaxation phase executes Dijkstra's algorithm on the core graph using $\tau_{\text{arr}}(\cdot, i)$ as tentative distances, with the priority queue initialized from all marked stops; any stop settled by the search is subsequently marked. Since Dijkstra on the core graph only guarantees shortest paths between stop pairs, initial and final transfers for non-stop source and target vertices $s, t \in V$ are handled separately via searches on the upward and downward graphs produced by Core-CH, respectively.

Like RAPTOR, MR correctly handles buffer times because it inherits RAPTOR's trip-aware route scanning phase.

\subsubsection*{ULTRA}

ULTRA (UnLimited TRAnsfers)~\cite{baum2019ultra} is a preprocessing technique that enables efficient multimodal journey planning with unrestricted transfers. Given an unlimited transfer graph representing any non-schedule-based transportation mode (e.g., walking, cycling), ULTRA computes a small set of transfer shortcuts that are provably sufficient for finding Pareto-optimal journeys with respect to arrival time and number of trips.

At query time, transfers from the source and to the target are computed using Bucket-CH one-to-many searches, while the precomputed shortcuts handle transfers between trips. This approach can be integrated with various public transit algorithms, including RAPTOR, CSA, and Trip-Based Routing, enabling unlimited transfers without sacrificing query performance.

\subsubsection*{CSA}

The Connection Scan Algorithm (CSA)~\cite{csa} is a simple yet efficient approach for answering earliest arrival queries in public transit networks. Unlike graph-based methods, CSA operates directly on the timetable by scanning elementary connections in chronological order.

A \emph{connection} $c = (v_{\text{dep}}, v_{\text{arr}}, \tau_{\text{dep}}, \tau_{\text{arr}}, T)$ represents a vehicle traveling from departure stop $v_{\text{dep}}$ to arrival stop $v_{\text{arr}}$, departing at time $\tau_{\text{dep}}$ and arriving at time $\tau_{\text{arr}}$, as part of trip $T$. CSA maintains a tentative arrival time $\tau_{\text{arr}}(v)$ for each stop $v$, initialized to $\infty$ except for the source stop $s$ which is set to the query departure time. Connections are stored in a single array sorted by $\tau_{\text{dep}}$. The algorithm scans this array sequentially: a connection $c$ is relaxed if the departure stop $v_{\text{dep}}$ is reachable in time, i.e., $\tau_{\text{arr}}(v_{\text{dep}}) \leq \tau_{\text{dep}}(c)$. When relaxed, the tentative arrival time at the stop is updated if it improves. The scan terminates once connections depart after the best known arrival time at the target.

CSA handles buffer times by tracking which trip (if any) was used to reach each stop. A passenger arriving on trip $T$ can board another connection of $T$ without waiting, but must respect the buffer time when boarding a different trip.

CSA's simplicity yields excellent cache locality and predictable memory access patterns. When combined with ULTRA preprocessing (ULTRA-CSA), it achieves state-of-the-art query performance for one-to-one earliest arrival queries with unlimited transfers~\cite{baum2019ultra}. We compare our approach against ULTRA-CSA as it represents the fastest known algorithm for this problem setting.

\section{Transfer Aware Dijkstra (TAD)}

TAD is a modification of TD-Dijkstra that correctly handles buffer times without requiring dominated connection filtering. The key insight is that when boarding a trip, TAD scans the entire remaining trip sequence rather than processing individual edges. This allows passengers to remain seated through intermediate stops without incurring buffer times, matching the behavior of timetable-based algorithms like CSA and RAPTOR.

The role of trip scanning in TAD is different from its role in timetable-based algorithms. RAPTOR-based multimodal algorithms organize the search in rounds and relax transfers from many marked stops at once, while CSA scans connections globally in chronological order. TAD keeps the single-source label-setting structure of Dijkstra: transfers are relaxed incrementally on the graph, and trip sequences are scanned only when a settled stop makes them reachable. TAD runs on the core graph, and its performance can be accelerated using Bucket-CH, as with TD-Dijkstra. 

\subsection{Algorithm Description}

TAD maintains node labels $\tau_{\text{arr}}[v]$ storing the earliest arrival time at each vertex $v \in V$. Algorithm~\ref{alg:tad} presents the main loop. For each settled vertex $u$, TAD examines all outgoing edges. For edges representing public transit connections, it identifies all trips departing from $u$ after the current arrival time plus the buffer time. For each such trip, the \textsc{ScanTrip} procedure (Algorithm~\ref{alg:scantrip}) processes the entire remaining trip sequence, updating node labels at each subsequent stop.

\subsubsection*{Handling Buffer Times}

When a vertex $u$ is settled with arrival time $\tau_{\text{arr}}[u]$, TAD determines the earliest time at which the passenger can board a departing trip by adding the buffer time: the earliest boarding time is $\tau_{\text{arr}}[u] + \beta(u)$.

Once a trip is boarded, \textsc{ScanTrip} processes all subsequent stops using the arrival times from the timetable. The passenger remains seated throughout, so no buffer times are incurred at intermediate stops. This correctly models the real-world behavior where seated passengers may continue without waiting.

\subsubsection*{Trip Pruning}

To handle non-FIFO schedules, TAD must scan multiple trips departing from the same stop, presorted by departure time, since a later-departing trip may arrive earlier at the final destination. However, scanning all trips is inefficient. For a fixed transit edge $e = (u, v)$, let $\tau_{\text{min}}(T, u)$ denote the minimum arrival time at $v$ among trip $T$ and all later trips in the departure-time-sorted list from $u$ to $v$; this value is stored as a suffix minimum for the trip list. Trip pruning terminates the iteration early when no remaining trip can improve over the current best arrival time, accounting for the buffer time at the destination stop $v$. Specifically, if $\tau_{\text{min}}(T, u) > \tau_{\text{best}} + \beta(v)$, all remaining trips can be skipped. 

Since $\beta(v)$ is a property of the timetable, the +$\beta(v)$ term in the cutoff accounts for the buffer wait incurred by transferring passengers at $v$. A label arriving at $v$ at time $\tau_{\text{best}}$ can board another trip only after waiting until $\tau_{\text{best}} + \beta(v)$, whereas a passenger already on a trip stays on board without waiting. After $\tau_{\text{best}} + \beta(v)$, the best known arrival at $v$ can board any remaining trip, so scanning those trips from $u$ is unnecessary. Larger buffer times widen this interval and can therefore reduce pruning effectiveness.

Consider trips $T_1, T_2, T_3, T_4$ departing from $A$ to $B$ (with $\beta(B) = 0$) with arrivals 9:30, 9:00, 10:00, 9:30 respectively, and $\tau_{\text{min}}$ values (minimum arrival among remaining trips) of 9:00, 9:00, 9:30, 9:30. With pruning (and $\beta(B) = 0$):
\begin{enumerate}
    \item Scan $T_1$ $\to$ $\tau_{\text{best}} = \text{09:30}$
    \item Scan $T_2$ $\to$ $\tau_{\text{best}} = \text{09:00}$ (improved)
    \item Check $T_3$: $\tau_{\text{min}}(T_3) = \text{09:30} > \text{09:00} + 0$ $\to$ \textbf{break}
\end{enumerate}
Trips $T_3$ and $T_4$ are skipped because their minimum arrival exceeds $\tau_{\text{best}} + \beta(B)$. When $\beta(B) > 0$, the pruning threshold is relaxed, allowing more trips to be scanned to account for the buffer wait at the destination.

\begin{algorithm}[t]
\small
\caption{Transfer Aware Dijkstra (TAD)}
\label{alg:tad}
\KwIn{Graph $G$, source $s$, departure time $\tau_{\text{dep}}$, target $t$}
\KwOut{Earliest arrival time at $t$}
$\tau_{\text{arr}}[v] \gets \infty$ for all $v \in V$\;
$\tau_{\text{arr}}[s] \gets \tau_{\text{dep}}$\;
$Q \gets \{s\}$ \tcp*{Priority queue keyed by $\tau_{\text{arr}}$}
\While{$Q \neq \emptyset$}{
    $u \gets Q.\text{ExtractMin}()$\;
    \lIf{$u = t$}{\textbf{break}}
    \ForEach{edge $e = (u, v) \in E$}{
        \If{$e$ is a transit edge \textbf{and} $u \in \mathcal{S}$}{
            $\tau_{\text{board}} \gets \tau_{\text{arr}}[u] + \beta(u)$ \tcp*{Add buffer time}
            $\tau_{\text{best}} \gets \infty$\;
            \ForEach{trip $T$ with $\tau_{\text{dep}}(T, u) \geq \tau_{\text{board}}$}{
                \If{$\tau_{\text{best}} \neq \infty$ \textbf{and} $\tau_{\text{min}}(T, u) > \tau_{\text{best}} + \beta(v)$}{
                    \textbf{break} \tcp*{Trip pruning}
                }

                $i \gets$ index of $u$ in $T$\;
                \textsc{ScanTrip}($T$, $i$)\;
                $\tau_{\text{best}} \gets \min(\tau_{\text{best}}, \tau_{\text{arr}}(T[i+1]))$\;
            
            }
        }
        \If{$e$ is a walking edge}{
            $\tau' \gets \tau_{\text{arr}}[u] + \tau_{\text{tra}}(e)$\;
            \If{$\tau' < \tau_{\text{arr}}[v]$}{
                $\tau_{\text{arr}}[v] \gets \tau'$\;
                $Q.\text{Update}(v)$\;
            }
        }
    }
}
\Return $\tau_{\text{arr}}[t]$\;
\end{algorithm}

\begin{algorithm}[t]
\small
\caption{\textsc{ScanTrip}: Process remaining stops on trip $T$}
\label{alg:scantrip}
\KwIn{Trip $T$, start index $i_{\text{start}}$}
\For{$i \gets i_{\text{start}} + 1$ \KwTo $|T|$}{
    $v \gets v(T[i])$ \tcp*{Stop vertex at index $i$}
    $\tau_{\text{cur}} \gets \tau_{\text{arr}}(T[i])$ \tcp*{Arrival time from timetable}
    \If{$\tau_{\text{cur}} < \tau_{\text{arr}}[v]$}{
        $\tau_{\text{arr}}[v] \gets \tau_{\text{cur}}$\;
        $Q.\text{Update}(v)$\;
    }
}
\end{algorithm}

\subsubsection*{Implementation Note}
The pseudocode in Algorithm~\ref{alg:tad} adds the buffer time at query time ($\tau_{\text{board}} \gets \tau_{\text{arr}}[u] + \beta(u)$) for clarity. Our implementation uses an equivalent formulation: during graph construction, each departure time is pre-decremented by the buffer time of the departure stop, i.e., stored as $\tau_{\text{dep}}(T, u) - \beta(u)$. The boarding condition $\tau_{\text{dep}}(T, u) \geq \tau_{\text{arr}}[u] + \beta(u)$ then simplifies to $(\tau_{\text{dep}}(T, u) - \beta(u)) \geq \tau_{\text{arr}}[u]$, eliminating an addition from the inner loop. This pre-subtraction does not affect \textsc{ScanTrip}, which uses unmodified timetable arrival times stored in the trip leg array.

\subsection{Correctness with Buffer Times}

\textsc{ScanTrip} correctly handles buffer times by processing entire trips rather than individual edges. Applied to the example from Section~II with $\beta(B) = 20$~min, TAD settles $A$ at 07{:}50, and \textsc{ScanTrip} on $T_1$ writes $\tau_{\text{arr}}[C] = 10{:}30$ before $B$ is settled. When $B$ later settles at 09{:}30, the buffer prevents boarding $T_1$ at 09{:}40, but $C$ has already been reached. TD-Dijkstra with filtering, by contrast, would have removed $T_1$'s $A \to B$ connection and missed this journey entirely.

The motivating example illustrates the algorithm on one input; the same property holds in general. Let $\tau^*(v)$ denote the earliest arrival time at $v$ over all valid journeys from source $s$ to $v$, under the standard assumption of non-negative transfer times and buffers and strictly positive trip-segment durations. We claim the invariant: when TAD extracts $v$ from the priority queue, $\tau_{\text{arr}}[v] = \tau^*(v)$. The base case is immediate: $s$ is extracted first with $\tau_{\text{arr}}[s] = \tau_{\text{dep}} = \tau^*(s)$. For the inductive step, take any optimal $s$--$v$ journey $J^*$ and decompose it by its last move. If $J^*$ ends with a transfer edge $(w, v)$, then $\tau^*(w) \leq \tau^*(v)$, so $w$ settles before $v$ with $\tau_{\text{arr}}[w] = \tau^*(w)$ by induction; relaxing $(w, v)$ when $w$ was processed gave $\tau_{\text{arr}}[v] \leq \tau^*(v)$. If $J^*$ ends with a trip segment $T$ from boarding stop $u$ to $v$ (with index $i$ at $u$, index $j$ at $v$, $j > i$), the boarding condition requires $\tau^*(u) + \beta(u) \leq \tau_{\text{dep}}(T[i]) \leq \tau_{\text{arr}}(T[j]) = \tau^*(v)$, so $u$ settles before $v$ with $\tau_{\text{arr}}[u] = \tau^*(u)$; \textsc{ScanTrip}$(T, i)$ at $u$ then writes $\tau_{\text{arr}}[v] \leq \tau_{\text{arr}}(T[j]) = \tau^*(v)$. The algorithm models the seated/transferring asymmetry through this structure: \textsc{ScanTrip} writes downstream arrivals directly from the timetable (bypassing $\beta$), while the +$\beta(u)$ term appears only in the boarding condition. Trip pruning preserves this invariant by the argument given in Section~III-A. Termination follows because each vertex is settled at most once; when $t$ is extracted, $\tau_{\text{arr}}[t] = \tau^*(t)$.

\section{Experiments}

\subsubsection{Experimental Setup}

We used the original implementation of MR, Core-CH, Bucket-CH, and ULTRA-CSA from the Karlsruhe Institute of Technology.\footnote{\url{https://github.com/kit-algo/ULTRA}} We extended this codebase with our TAD implementation and made the fork publicly available.\footnote{\url{https://github.com/andrii-rohovyi/PublicTransitRoutingWithUnlimitedTransfer}}

All code was compiled using the GNU C++ Compiler (g++) version 13.3.0. Experiments were performed on an AMD Ryzen 9 7900X (12 cores, 24 threads) running Ubuntu 24.04.3 LTS under WSL2 (Linux kernel 6.6.87.2-microsoft-standard-WSL2) on x86-64. The WSL2 VM had 30~GiB RAM available (host system: 64~GiB).

\subsubsection{Datasets}

We evaluated our approach on the London and Switzerland networks, which have been used in previous unlimited transfer research from KIT~\cite{baum2019ultra, ultra_mcraptor}.\footnote{\url{https://i11www.iti.kit.edu/PublicTransitData/ULTRA/}} Table~\ref{tab:datasets} summarizes the network characteristics.

\begin{table}[t]
\centering
\caption{Dataset characteristics.}
\label{tab:datasets}
\begin{tabular}{lrr}
& London & Switzerland \\
\midrule
Stops & 19,682 & 25,125 \\
Trips & 114,508 & 350,006 \\
Routes & 1,955 & 13,786 \\
Connections & 4,394,136 & 4,336,859 \\
Transfer graph vertices & 181,642 & 603,691 \\
Transfer graph edges & 334,112 & 465,067 \\
Stops with buffer $> 0$ & 0 & 12,847 \\
\end{tabular}
\end{table}

The unrestricted transfer graphs for both networks were constructed following the methodology described in~\cite{baum2019ultra}. Road graphs, including pedestrian zones and staircases, were extracted from OpenStreetMap.\footnote{\url{https://download.geofabrik.de/}} Walking was used as the transfer mode with a constant speed of 4.5~km/h. The transfer graph was connected to the public transit network by matching stops to their nearest vertices in the road graph, with vertices of degree one or two contracted unless they coincided with stops.

Importantly, these two datasets exhibit different characteristics with respect to buffer times. The London network has no buffer times: all stops have $\beta(v) = 0$. This allows us to compare all algorithms, including TD-Dijkstra with dominated connection filtering. The Switzerland network has buffer times at 12,847 stops (51\% of all stops), drawn from same-stop entries in the SBB / Open Data CH GTFS \texttt{transfers.txt}; most values fall between 1 and 10 minutes. Since dominated connection filtering breaks down with buffer times, we exclude TD-Dijkstra and TTN variants from the Switzerland evaluation and compare only algorithms that correctly handle buffer times: MR, TAD, and ULTRA-CSA.

\subsubsection{Results}

Table~\ref{tab:results} reports average query times over 1{,}000 random source-target pairs, with both endpoints drawn uniformly from the transfer graph vertices and departure times drawn uniformly from a 24-hour window. The same fixed-seed query set is used across algorithms, and wall-clock timings exclude data loading and preprocessing. Dashes mark algorithms that cannot be applied due to buffer-time incompatibility. 

\begin{table}[t]
\centering
\caption{Average query times [ms].}
\label{tab:results}
\begin{tabular}{lrrrr}
& \multicolumn{2}{c}{London} & \multicolumn{2}{c}{Switzerland} \\
\cmidrule(lr){2-3}\cmidrule(lr){4-5}
Algorithm & Time & Speedup & Time & Speedup \\
\midrule
MR             & 12.85 & 1.00$\times$ & 26.12 & 1.00$\times$ \\
TD-Dijkstra (no  Bucket-CH)   &  8.03 & 1.60$\times$ & ---   & --- \\
TD-Dijkstra (Bucket-CH)  &  4.47 & 2.88$\times$ & ---   & --- \\
TAD  (no  Bucket-CH)         &  9.73 & 1.32$\times$ & 16.97 & 1.54$\times$ \\
TAD (Bucket-CH)          &  5.93 & 2.17$\times$ &  9.08 & 2.88$\times$ \\
ULTRA-CSA                &  1.64 & 7.86$\times$ &  2.40 & 10.88$\times$ \\
\end{tabular}
\end{table}

\begin{table}[t]
\centering
\caption{TTN variants vs TD-Dijkstra on London.}
\label{tab:ttn_comparison}
\begin{tabular}{lrr}
Algorithm & Time [ms] & vs TD-Dijkstra \\
\midrule
TD-Dijkstra (no  Bucket-CH)& 8.03 & 1.00$\times$ \\
TTN-FC (no  Bucket-CH)& 10.31 & 0.78$\times$ \\
TTN-CST  (no  Bucket-CH)& 10.46 & 0.77$\times$ \\
TTN-BST  (no  Bucket-CH)& 15.22 & 0.53$\times$ \\
\midrule
TD-Dijkstra (Bucket-CH) & 4.47 & 1.00$\times$ \\
TTN-FC (Bucket-CH) & 6.16 & 0.73$\times$ \\
TTN-CST (Bucket-CH) & 6.47 & 0.69$\times$ \\
TTN-BST (Bucket-CH) & 10.82 & 0.41$\times$ \\
\end{tabular}
\end{table}

We compare against the original MR, which has no Bucket-CH acceleration, so we also evaluate TD-Dijkstra and TAD without it for a direct comparison. Even without Bucket-CH, both are faster than MR: 1.32x (TAD) and 1.60x (TD-Dijkstra) on London, and 1.54x for TAD on Switzerland.

On London (no buffer times), all algorithms produce identical results. TD-Dijkstra with Bucket-CH is the fastest among algorithms without ULTRA preprocessing, reaching a 2.88$\times$ speedup over MR; TAD's trip-scanning overhead reduces this to 2.17$\times$. Bucket-CH significantly accelerates TD-Dijkstra search. ULTRA-CSA remains the fastest overall at 7.86$\times$ via its precomputed transfer shortcuts.

Table~\ref{tab:ttn_comparison} compares TTN variants against TD-Dijkstra on London. All six are slower than the per-edge baseline---28--90\% with Core-CH and 38--142\% with Bucket-CH. TTN-BST more than doubles query time with Bucket-CH. Two factors explain this: contiguous \texttt{std::vector} storage already provides excellent cache locality for sequential binary searches, and the auxiliary structures required by TTN (cascading pointers, combined trees) introduce indirection that outweighs the reduction in search count. TTN cannot be applied on Switzerland, since it inherits TD-Dijkstra's reliance on dominated connection filtering, which is unsound with buffer times.

On Switzerland (51\% of stops have $\beta(v) > 0$), TD-Dijkstra and TTN are inapplicable. Among the algorithms that correctly handle buffers, TAD with Bucket-CH achieves a 2.88$\times$ speedup over MR---the fastest correct algorithm without ULTRA preprocessing. ULTRA-CSA reaches 10.88$\times$ via precomputed shortcuts, with the trade-off discussed in Section~\ref{sec:preprocessing}.

\subsubsection{Preprocessing}
\label{sec:preprocessing}

Table~\ref{tab:preprocessing} reports preprocessing times for the speedup techniques. Following prior work~\cite{baum2019ultra, sauer2024closing}, we omit shared pipeline costs (GTFS parsing, intermediate conversion) and data structure construction times (under 20\,s for all algorithms).

Bucket-CH and Core-CH construction are one-time costs for a fixed transfer graph. ULTRA stop-to-stop shortcuts dominate preprocessing at 7--14 minutes and must be recomputed when delay assumptions change, whereas TAD operates directly on the timetable without precomputed shortcuts.

\begin{table}[t]
\centering
\caption{Preprocessing times for speedup techniques.}
\label{tab:preprocessing}
\begin{tabular}{lrr}
Component & London & Switzerland \\
\midrule
Core-CH construction & 1.81\,s & 3.66\,s \\
CH construction (Bucket-CH) & 7.25\,s & 33.37\,s \\
ULTRA stop-to-stop shortcuts & 13\,m\,41\,s & 7\,m\,2\,s \\
\end{tabular}
\end{table}

These times clarify the deployment trade-off between TAD and ULTRA-CSA. For static timetables and high query volumes, ULTRA-CSA is the preferred choice among the evaluated algorithms for one-to-one earliest-arrival queries. Relative to TAD with Bucket-CH, the ULTRA shortcut preprocessing in Table~\ref{tab:preprocessing} is amortized after roughly 191,000 London queries and 63,000 Switzerland queries. Below these query volumes, or when timetable or delay assumptions change frequently enough that shortcuts must be recomputed, TAD avoids the minutes-long ULTRA shortcut phase and remains a practical choice despite slower individual queries.

\section{Conclusion and Future Steps}

We have demonstrated that classical Dijkstra-based approaches can outperform MR for the unlimited transfer problem in public transit routing. However, we identified a critical limitation: efficient TD-Dijkstra implementations rely on dominated connection filtering, which is invalid when stops have buffer times.

The fundamental issue is that dominated connection filtering analyzes edges in isolation, without distinguishing between seated passengers (who may continue without waiting) and transferring passengers (who must respect buffer times). A connection that appears dominated on a single edge may be part of a longer trip where remaining seated provides a faster overall journey.

To address this limitation, we introduced Transfer Aware Dijkstra (TAD), a modification of TD-Dijkstra that scans entire trip sequences rather than individual edges. When a passenger boards a trip, TAD processes all subsequent stops, correctly allowing the passenger to remain seated without incurring buffer times at intermediate stops.

Our experiments on the London and Switzerland networks reveal important findings:

\begin{itemize}
\item On data without buffer times, TD-Dijkstra with Bucket-CH achieves a 2.88$\times$ speedup over MR, making it the fastest algorithm for unlimited transfers without ULTRA preprocessing. TAD achieves a 2.17$\times$ speedup with slightly more overhead due to trip scanning.
\item On data with buffer times, TD-Dijkstra cannot be applied due to invalid dominated connection filtering. TAD with Bucket-CH correctly handles buffer times and achieves a 2.88$\times$ speedup over MR.
\item TAD provides a robust solution that works correctly on both networks, combining correctness with performance advantages over MR.
\item Timetable Nodes (TTN), which provides 10--30\% speedups in Python, is 28--142\% slower in C++. The auxiliary data structures required by TTN introduce memory indirection that outweighs the reduction in binary searches, given the excellent cache locality of contiguous vectors and effective CPU prefetching already present in the baseline.
\end{itemize}

ULTRA-CSA remains the fastest algorithm overall, but requires significant preprocessing time that may be impractical when timetables change frequently. Recent work on delay-robust multimodal journey planning~\cite{sauer2024closing, bez2024delay} has addressed this limitation for ULTRA with Trip-Based routing, though it requires assumptions about maximum delay bounds. TAD does not require such assumptions, as it operates directly on the timetable without precomputed shortcuts and naturally supports real-time delay updates.

Several extensions remain for future work. Walking speeds in our evaluation are a constant 4.5~km/h, but in practice they vary between users. TAD can support this by storing distances on transfer edges and dividing by a per-query speed during relaxation time, as has been done in previous work~\cite {rohovyi2025multimodal, abuaisha2024practical}. Bicriteria extensions of TAD, such as jointly minimising arrival time and transfer count, can be explored.

\IEEEtriggeratref{19}
\IEEEtriggercmd{\enlargethispage{2\baselineskip}}
\bibliographystyle{IEEEtran}
\bibliography{bibliography}

\end{document}